\documentclass[10pt,conference]{IEEEtran}
\IEEEoverridecommandlockouts

\usepackage{cite}
\usepackage{amsmath,amssymb,amsfonts}
\usepackage{algorithmic}
\usepackage{graphicx}
\usepackage{textcomp}
\usepackage{xcolor}
\usepackage{subcaption}
\usepackage{braket}
\usepackage{tikz}
\usepackage{qcircuit}
\usepackage{booktabs} 
\usepackage{tabularx}
\usepackage{caption}
\usepackage{longtable}

\def\BibTeX{{\rm B\kern-.05em{\sc i\kern-.025em b}\kern-.08em
T\kern-.1667em\lower.7ex\hbox{E}\kern-.125emX}}

\begin{document}

\title{Benchmarking Classical and Quantum Models for DeFi Yield Prediction on Curve Finance \thanks{The views expressed in this article are those of the authors and do not represent the views of Omnis Labs. This article is for informational purposes only. Nothing contained in this article should be construed as investment advice. Omnis Labs makes no express or implied warranties and expressly disclaims all legal, tax, and accounting implications related to this article.}}

\author{
\IEEEauthorblockN{Chi-Sheng Chen}
\IEEEauthorblockA{
% \textit{Neuro Industry Research} \\
\textit{Omnis Labs}\\
% Cambridge, Massachusetts, USA \\
michael@omnis.farm}
\and
\IEEEauthorblockN{Aidan Hung-Wen Tsai}
\IEEEauthorblockA{\textit{Omnis Labs} \\
% New York, USA \\
aidan@omnis.farm}
% \and
% \IEEEauthorblockN{Yun-Cheng Tsai}
% \IEEEauthorblockA{\textit{National Taiwan Normal University}\\
% Taipei, Taiwan \\
% pecu@ntnu.edu.tw}
}

\maketitle

\begin{abstract}
The rise of decentralized finance (DeFi) has created a growing demand for accurate yield and performance forecasting to guide liquidity allocation strategies. In this study, we benchmark six models—XGBoost, Random Forest, LSTM, Transformer, quantum neural networks (QNN), and quantum support vector machines with quantum feature maps (QSVM-QNN)—on one year of historical data from 28 Curve Finance pools. We evaluate model performance on test MAE, RMSE, and directional accuracy. Our results show that classical ensemble models, particularly XGBoost and Random Forest, consistently outperform both deep learning and quantum models. XGBoost achieves the highest directional accuracy (71.57\%) with a test MAE of 1.80, while Random Forest attains the lowest test MAE of 1.77 and 71.36\% accuracy. In contrast, quantum models underperform with directional accuracy below 50\% and higher errors, highlighting current limitations in applying quantum machine learning to real-world DeFi time series data. This work offers a reproducible benchmark and practical insights into model suitability for DeFi applications, emphasizing the robustness of classical methods over emerging quantum approaches in this domain.
\end{abstract}


\section{Introduction}
The decentralized finance (DeFi) ecosystem has rapidly emerged as a cornerstone of the blockchain economy, facilitating billions of dollars in on-chain liquidity, lending, and trading without intermediaries. Within this ecosystem, protocols such as Curve Finance play a vital role in optimizing stable asset swaps and yield farming through algorithmically managed liquidity pools. Accurately forecasting the dynamics of these pools, such as yield changes, total value locked (TVL), or trade volume, is crucial for designing profitable trading strategies, allocating capital, and mitigating risk in DeFi applications.

While classical time series models and deep learning approaches have been extensively applied to traditional finance, their utility in the DeFi domain remains underexplored. Moreover, recent advancements in quantum machine learning (QML) suggest that quantum-enhanced models may offer advantages in learning complex financial patterns, particularly under limited data or non-convex optimization scenarios. However, the practical efficacy of QML models in real-world blockchain settings is still uncertain and lacks empirical benchmarking.

In this work, we present a comprehensive benchmark study of six machine learning models on a newly curated dataset derived from 28 Curve Finance~\cite{curve_stablecoin_whitepaper} pools over a full year. We compare two classical ensemble models (XGBoost~\cite{chen2016xgboost}, Random Forest~\cite{breiman2001random}), two deep learning models (LSTM~\cite{hochreiter1997lstm}, Transformer~\cite{vaswani2017attention}), and two quantum models (QNN~\cite{schuld2014quest}, QSVM-QNN~\cite{rebentrost2014quantum} with quantum feature maps), all trained to predict future yield-related values. Each model is evaluated using standard forecasting metrics: Mean Absolute Error (MAE), Root Mean Squared Error (RMSE), and directional accuracy.

Our findings highlight that traditional ensemble methods continue to dominate in this DeFi forecasting task. Notably, XGBoost and Random Forest consistently outperform neural and quantum models in both pointwise prediction and directional accuracy. Conversely, quantum models, despite theoretical potential, struggle with generalization and noise sensitivity in this context.

The contributions of this paper are threefold:
\begin{itemize}
    \item We construct a standardized DeFi time-series forecasting benchmark using real-world Curve Finance data.
    \item We provide a unified evaluation of classical, deep learning, and quantum models under consistent experimental settings.
    \item We offer practical insights into the strengths and limitations of quantum models in the context of financial time-series prediction.
\end{itemize}

This study provides a reproducible foundation for future work in quantum finance, DeFi forecasting, and cross-paradigm machine learning comparisons.

\section{Related Work}
\subsection{Machine Learning for Cryptocurrency and DeFi Forecasting}
Early studies on on‑chain analytics mainly focused on price prediction of major cryptocurrencies using classical statistical techniques. With the advent of gradient‑boosting ensembles, XGBoost became a popular baseline due to its capacity to capture non‑linear feature interactions. Academic work that explicitly targets DeFi liquidity pools remains sparse; most existing analyses appear as industry white papers or blogs. This gap underlines the need for systematic benchmarks on real‑world DeFi datasets such as Curve Finance.

\subsection{Deep Learning for Financial Time-Series and Multimodal Forecasting}

Financial forecasting has long been a prominent domain for the application of machine learning, owing to the high complexity of financial dynamics and the substantial impact that even marginal predictive improvements can yield. While traditional approaches often rely on structured numerical indicators extracted from financial statements, recent advances have expanded into the integration of unstructured data sources, such as earnings call transcripts and audio recordings, for enhanced modeling capacity.

In this context, deep learning models have been proposed to better incorporate the semantic and quantitative features present in such multimodal financial data. A recent line of work~\cite{yang2022numhtml} introduces a numeric-aware hierarchical transformer architecture that explicitly distinguishes between numerical categories (e.g., monetary values, percentages, temporal indicators) and leverages their magnitude in prediction tasks such as return forecasting and risk estimation. These models align textual and numeric modalities to extract richer representations and improve generalization.

Empirical evaluations demonstrate that such architectures substantially outperform baseline models across several financial prediction benchmarks. Nevertheless, the practical application of these models in decentralized finance (DeFi) remains challenged by limited and noisy pool-specific data histories, which can hinder the training of data-hungry deep networks without strong regularization or transfer learning mechanisms.

\subsection{Quantum Machine Learning in Finance}

Quantum Machine Learning (QML) explores the integration of quantum computing principles into traditional machine learning pipelines, aiming to harness advantages such as Hilbert-space expressivity and quantum parallelism. In the context of financial modeling, QML has been proposed as a potential solution to high-dimensional and noisy data environments commonly found in forecasting and classification tasks~\cite{schaden2002quantum}. Preliminary studies have explored the use of parameterized quantum circuits and hybrid quantum-classical architectures for time-series prediction and asset classification. These approaches typically operate on small datasets or low-dimensional feature spaces due to current hardware limitations. While theoretical work suggests that quantum kernels and variational circuits may offer expressive power beyond classical models, empirical evidence under realistic noise conditions remains limited. As such, classical machine learning models, especially ensemble methods, continue to outperform QML counterparts in most large-scale financial applications~\cite{zhou2025quantum}. Nonetheless, the ongoing development of near-term quantum devices and improved noise mitigation techniques may gradually close this performance gap.

\subsection{Research Gap}
Existing literature provides rich evidence on classical and deep‑learning approaches for asset‑price and yield‑curve prediction, and an emerging body on QML for generic financial tasks.  
However, no prior work offers a head‑to‑head comparison of classical, deep, and quantum models on \emph{DeFi liquidity‑pool} data using consistent metrics.  
Our study fills this gap by benchmarking six representative models on 37 Curve Finance pools, revealing that ensemble tree methods (Random Forest, XGBoost) remain state‑of‑the‑art for DeFi yield forecasting, while current QML approaches lag behind under practical constraints.

% ------------------------------
% Example BibTeX keys referenced above (user should include full entries in .bib file)
% \bibitem{hybrid_lstm_xgb2025} ...
% \bibitem{multiyield_transformer2024} ...
% \bibitem{lstm_laglasso2024} ...
% \bibitem{lstm_extrapolation2024} ...
% \bibitem{qml_ts_forecast2022} ...
% \bibitem{qsvm_finance2025} ...
% \bibitem{quantum_fin_forecast2024} ...

\begin{figure}
    \centering
    \includegraphics[width=1\linewidth]{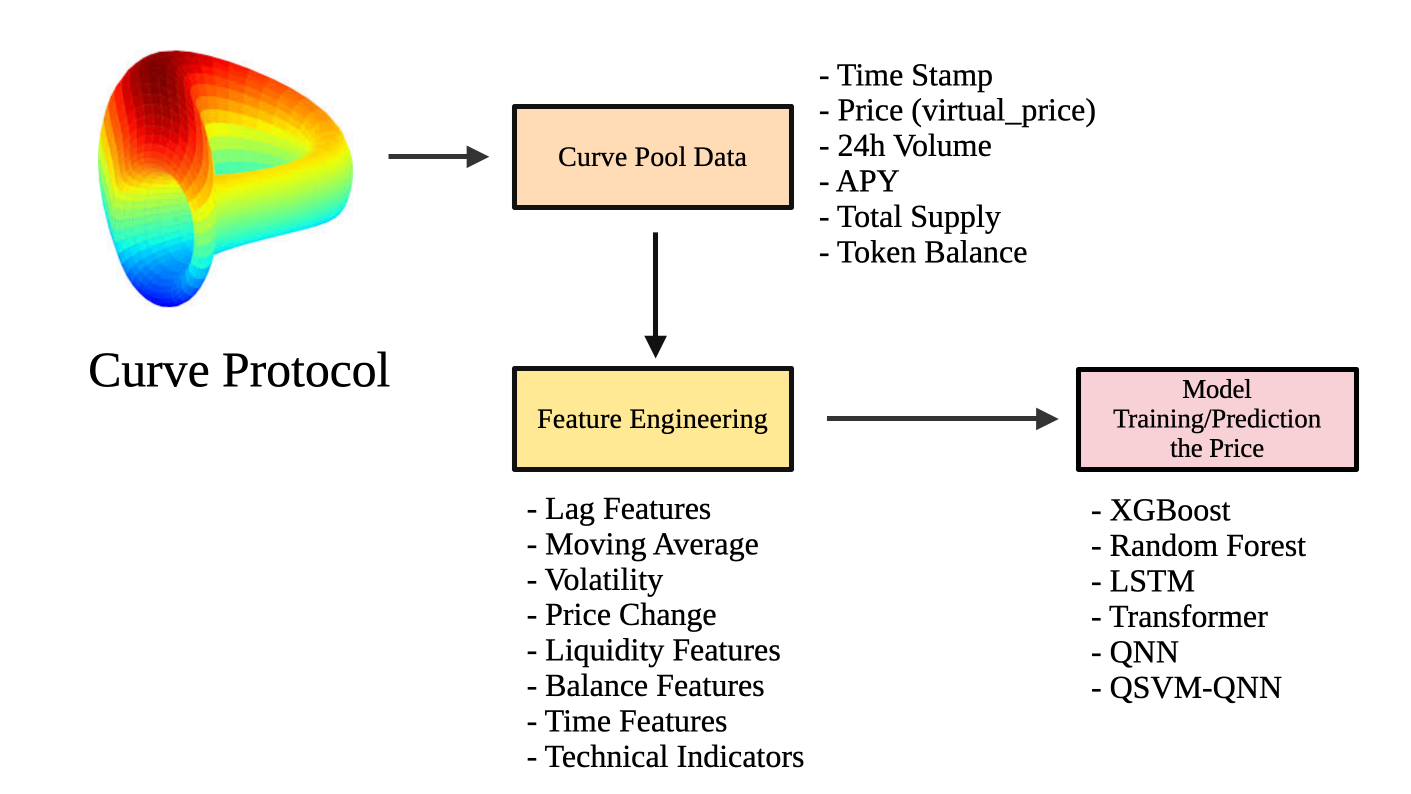}
    \caption{Overview of this work.}
    \label{fig:enter-label}
\end{figure}
\section{Methodology}
\subsection{Data Acquisition and Temporal Hold‑Out}  \label{ssec:data}
We curate a daily time–series data set for the past $365$ days\footnote{Snapshot date: \textit{T\textsubscript{end}} = \textit{2025‑07‑21}.} from all $P=28$ public liquidity pools on \textsc{Curve Finance}.
For each pool $p$, observations are ordered chronologically and partitioned into an in‑sample set $\mathcal D^{(p)}_\text{train}$ (first $80\%$) and an out‑of‑sample set $\mathcal D^{(p)}_\text{test}$ (last $20\%$):
\begin{equation}
\mathcal D^{(p)}
=\bigl\{\bigl(\mathbf x_{t}^{(p)}, \,y_{t+1}^{(p)}\bigr)\bigr\}_{t=1}^{T_p},
\quad 
\mathbf x_{t}^{(p)}\in\mathbb R^{d},\;
y_{t+1}^{(p)}\in\mathbb R,
\end{equation}
where $y_{t+1}^{(p)}$ denotes the next‑day closing price of pool $p$ at calendar day~$t+1$.

\subsection{Feature Engineering and Target Variable}\label{ssec:fe}

Our pipeline is \emph{multivariate}: beyond \texttt{virtual\_price}, we ingest liquidity balances, volume, APY, total supply, and derived ratios. Let
$\mathbf{r}_t \in \mathbb{R}^{d_0}$ denote the raw vector at time $t$
(\textit{e.g.}, virtual price, 24h volume, APY, total supply, token balances).
We transform $\mathbf{r}_t$ into a rich feature vector $\mathbf{x}_t \in \mathbb{R}^{d}$ via:

\begin{enumerate}
\item \textbf{Lag features (multi-scale).}
For a raw scalar series $z_t$, we create delayed copies at heterogeneous horizons
$\ell \in \{1,6,24,168\}$ (hours):
\begin{equation}
z^{(\ell)}_{t} = z_{t-\ell}.
\end{equation}

\item \textbf{Rolling statistics.}
For windows $k \in \{24,168,672\}$ (hours), we compute moving averages, standard deviations,
and coefficients of variation:
\begin{align}
\mathrm{MA}^{(k)}_t &= \frac{1}{k}\sum_{i=0}^{k-1} z_{t-i}, \\
\mathrm{STD}^{(k)}_t &= \sqrt{\frac{1}{k}\sum_{i=0}^{k-1}\bigl(z_{t-i}-\mathrm{MA}^{(k)}_t\bigr)^2}, \\
\mathrm{CV}^{(k)}_t &= \frac{\mathrm{STD}^{(k)}_t}{\mathrm{MA}^{(k)}_t + \varepsilon},
\end{align}
where $\varepsilon$ prevents division by zero.

\item \textbf{Price‑change signals.}
Absolute and logarithmic changes:
\begin{align}
\Delta z_t &= z_t - z_{t-1}, \\
\Delta_{\log} z_t &= \log z_t - \log z_{t-1}.
\end{align}

\item \textbf{Liquidity and balance ratios.}
Given token balances $\{b^{(j)}_t\}$, we form pool‑internal structure metrics:
\begin{align}
\mathrm{balance\_ratio}_t &= \frac{\max_j b^{(j)}_t}{\sum_j b^{(j)}_t}, \\
\mathrm{balance\_imbalance}_t &= \max_j b^{(j)}_t - \min_j b^{(j)}_t,
\end{align}
and analogous terms for \texttt{total\_supply} (changes, moving averages).

\item \textbf{Technical indicators.}
A 14‑period Relative Strength Index (RSI):
\begin{equation}
\mathrm{RSI}_t = 100 \left/ \left(1 + \frac{\overline{\text{loss}}_t}{\overline{\text{gain}}_t}\right)\right..
\end{equation}

\item \textbf{Temporal encodings.}
Calendar features via sinusoidal embeddings:
\begin{align}
\sin_h &= \sin\!\left(2\pi \frac{h_t}{24}\right),\quad
\cos_h = \cos\!\left(2\pi \frac{h_t}{24}\right),\\
\sin_d &= \sin\!\left(2\pi \frac{d_t}{7}\right),\quad
\cos_d = \cos\!\left(2\pi \frac{d_t}{7}\right),\\
\sin_m &= \sin\!\left(2\pi \frac{m_t}{12}\right),\quad
\cos_m = \cos\!\left(2\pi \frac{m_t}{12}\right).
\end{align}
\end{enumerate}

\subsection{Model Families}  \label{ssec:models}
Six representative predictors are trained on the identical feature space:

\begin{itemize}
\item \textbf{Random Forest} (RF): $n_{\text{tree}}\!=\!150$, bootstrap sampling, Gini split.
\item \textbf{Extreme Gradient Boosting} (XGB): tree depth $\le6$, learning rate $\eta$ chosen by validation, early stopping on $\mathcal D_\text{train}$.
\item \textbf{LSTM}: two stacked LSTM layers ($h=64$) followed by a fully‑connected head.
\item \textbf{Transformer}: two‑layer encoder ($h=8$ heads, $d_{\text{model}}\!=\!128$) with position encoding.
\item \textbf{Variational Quantum Neural Network} (QNN): $N_q=4$ qubits, four‑layer parameterised quantum circuit with entangling CNOT topology, trained via the parameter‑shift rule.
\item \textbf{QSVM‑QNN Hybrid}: a quantum feature map feeding a variational classifier with hinge loss.
\end{itemize}

Classical models are executed on CPU/GPU; quantum variants are batched after compute resources are released.

\subsection{Training Objective and Procedure}  \label{ssec:train}
For each pool $p$ and model $m$, parameters $\boldsymbol\theta^{(m)}$ minimise the mean‑squared error (MSE)
\begin{equation}
\mathcal L^{(m,p)}(\boldsymbol\theta)
=\frac{1}{N_\text{train}^{(p)}}\sum_{(\mathbf x,y)\in\mathcal D^{(p)}_\text{train}}
\bigl(y-\hat y^{(m)}(\mathbf x;\boldsymbol\theta)\bigr)^2,
\end{equation}
using Adam (deep models) or default package optimisers (tree and quantum models).  
Training stops at $100$ epochs or when validation loss fails to decrease for $10$ consecutive epochs.

\subsection{Evaluation Metrics}  \label{ssec:metrics}
Performance is quantified on both splits via:
\begin{align}
\text{MAE} &= \frac1N\sum_{i=1}^{N}\lvert y_i-\hat y_i\rvert, \\
\text{RMSE}&= \sqrt{\frac1N\sum_{i=1}^{N}\bigl(y_i-\hat y_i\bigr)^2}, \\
\text{DA}  &= \frac1N\sum_{i=1}^{N}\mathbf 1\!\left[
\operatorname{sgn}\!\bigl(y_i-y_{i-1}\bigr)=
\operatorname{sgn}\!\bigl(\hat y_i-y_{i-1}\bigr)
\right].
\end{align}

\subsection{Statistical Aggregation and Ranking}  \label{ssec:ranking}
Each metric $\phi\in\{\text{MAE},\text{RMSE},\text{DA}\}$ is first recorded pool‑wise,
$\phi^{(m,p)}$, then aggregated
\(
\bar\phi^{(m)}=\frac{1}{P}\sum_{p=1}^{P}\phi^{(m,p)},\;
\sigma_\phi^{(m)}=\sqrt{\frac{1}{P-1}\sum_{p}(\phi^{(m,p)}-\bar\phi^{(m)})^{2}}.
\)
Models are ranked primarily by descending $\bar{\text{DA}}^{(m)}$ on $\mathcal D_\text{test}$; $\sigma$ is reported to reveal robustness across heterogeneous pools.

\paragraph{Summary.}
The pipeline guarantees that classical, deep, and quantum predictors are contrasted under identical data splits and feature spaces, thereby isolating algorithmic advantages from confounding factors such as information leakage or inconsistent preprocessing.

\section{Experiments}
\subsection{Dataset and Pre‑Processing}
We compiled historical time series from 28 Curve Finance liquidity pools, covering the period from 2024-07-20 15:31:05 to 2025-07-20 09:31:05, for a total span of 364 days 17 hours 59 minutes 59 seconds. The final dataset contains 1,460 observations sampled at an average interval of approximately 6 hours (5\,h 59\,min 59\,s), yielding a uniform six-hour cadence. This sampling frequency is sufficient to capture intra-day fluctuations typical of DeFi markets while maintaining an almost year-long horizon. The series is temporally complete with no missing timestamps on the six-hour grid, ensuring a clean foundation for downstream forecasting tasks.
We collect day‑level snapshots for 28 Curve Finance pools covering 365~days, 4 points per day.
For every pool, we split the time series chronologically into
80\% training and 20\% testing.
Feature engineering follows two steps implemented in pipeline:
(1) lagged windows (1–7\,days), technical indicators
(moving average, volatility, RSI, log‑returns) and
(2) seasonal sine/cosine encodings of calendar time.
All continuous features are standardised prior to modelling.

Columns such as \texttt{timestamp}, \texttt{pool\_address}, \texttt{pool\_name}, \texttt{source}, and the direct
target fields are excluded from $\mathbf{x}_t$ to prevent leakage. All continuous features are $z$‑score normalised using
statistics from $\mathcal{D}_{\text{train}}$ and reused for $\mathcal{D}_{\text{test}}$.

\paragraph{Target Variable.}
We predict the 24‑hour ahead virtual price (or its return). Concretely,
\begin{equation}
\texttt{target\_24h} = \texttt{virtual\_price}_{t+24},
\end{equation}
and the percentage return is defined as
\begin{equation}
\texttt{target\_return\_24h} =
\left(\frac{\texttt{target\_24h}}{\texttt{virtual\_price}_t} - 1\right)\times 100.
\end{equation}
Both $\texttt{virtual\_price}_t$ and forward targets are removed from the input feature set to avoid information leakage.

\subsection{Model Portfolio}

We benchmark six algorithmic families. Given features $\mathbf{x}_t$ and target $y_{t+1}$, each model $f_\phi$ minimises a supervised loss $\mathcal{L}$ (MAE/RMSE).

\paragraph{Random Forest (RF).}
An ensemble of $T$ CART trees $\{h_t\}_{t=1}^T$:
\begin{equation}
\hat{y} = \frac{1}{T}\sum_{t=1}^{T} h_t(\mathbf{x}), \qquad h_t:\mathbb{R}^d \rightarrow \mathbb{R}.
\end{equation}

\paragraph{XGBoost (XGB).}
Additive tree boosting:
\begin{equation}
\hat{y}^{(K)}(\mathbf{x}) = \sum_{k=1}^{K} f_k(\mathbf{x}), \quad f_k \in \mathcal{F},
\end{equation}
\begin{equation}
\mathcal{L} = \sum_{i} \ell\!\left(y_i,\hat{y}_i^{(K-1)} + f_K(\mathbf{x}_i)\right)
+ \Omega(f_K), 
\end{equation}
\begin{equation}
\Omega(f)=\gamma T_f + \frac{1}{2}\lambda \lVert \mathbf{w}_f\rVert_2^2.
\end{equation}

\paragraph{LSTM.}
For windowed inputs $\mathbf{X}_{t-L+1:t}$, the cell updates are:
\begin{equation}
\mathbf{i}_t = \sigma(W_i\mathbf{x}_t + U_i\mathbf{h}_{t-1}+ \mathbf{b}_i), \quad
\mathbf{f}_t = \sigma(W_f\mathbf{x}_t + U_f\mathbf{h}_{t-1}+ \mathbf{b}_f),
\end{equation}
\begin{equation}
\mathbf{o}_t = \sigma(W_o\mathbf{x}_t + U_o\mathbf{h}_{t-1}+ \mathbf{b}_o), \quad
\tilde{\mathbf{c}}_t = \tanh(W_c\mathbf{x}_t + U_c\mathbf{h}_{t-1}+ \mathbf{b}_c),
\end{equation}
\begin{equation}
\mathbf{c}_t = \mathbf{f}_t \odot \mathbf{c}_{t-1} + \mathbf{i}_t \odot \tilde{\mathbf{c}}_t,\qquad
\mathbf{h}_t = \mathbf{o}_t \odot \tanh(\mathbf{c}_t),
\end{equation}
\begin{equation}
\hat{y}=W^{\text{fc}}\mathbf{h}_t+b^{\text{fc}}.
\end{equation}

\paragraph{Transformer Encoder.}
Self-attention with queries/keys/values $\mathbf{Q},\mathbf{K},\mathbf{V}$:
\begin{equation}
\mathrm{Attn}(\mathbf{Q},\mathbf{K},\mathbf{V})=
\mathrm{softmax}\!\left(\frac{\mathbf{Q}\mathbf{K}^\top}{\sqrt{d_k}}\right)\mathbf{V}.
\end{equation}
Two encoder blocks (multi-head attention + FFN) produce a pooled vector projected to $\hat{y}$.

\paragraph{Quantum Neural Network (QNN).}
Classical $\mathbf{x}$ is encoded by $U_{\text{enc}}(\mathbf{x})$ on $n$ qubits, followed by a variational circuit $U_\theta$:
\begin{equation}
\ket{\psi(\mathbf{x};\theta)} = U_\theta\, U_{\text{enc}}(\mathbf{x})\, \ket{0}^{\otimes n},
\end{equation}
\begin{equation}
\hat{y} = \bra{\psi(\mathbf{x};\theta)} M \ket{\psi(\mathbf{x};\theta)},
\end{equation}
with $M$ a Pauli observable; gradients via parameter-shift.

\paragraph{QSVM–QNN.}
Quantum kernel:
\begin{equation}
K(\mathbf{x},\mathbf{x}') =
\left|\bra{0} U_{\text{enc}}^\dagger(\mathbf{x}) U_{\text{enc}}(\mathbf{x}') \ket{0}\right|^2,
\end{equation}
used inside an SVM-style predictor (regression variant with a variational head):
\begin{equation}
f(\mathbf{x}) = \sum_i \alpha_i\, K(\mathbf{x},\mathbf{x}_i) + b.
\end{equation}

% \subsection{Training Procedure and Automation}
% A controller class \verb|BatchModelComparison| orchestrates the
% end‑to‑end workflow (load, feature creation, training, evaluation, logging)
% for each dataset.
% Real‑time metrics are appended to \texttt{realtime\_results.csv} immediately after
% each model finishes, enabling progress monitoring and crash recovery
% (\verb|set_realtime_write(True)|).
% Intermediate backups are saved every five datasets, and a full
% \texttt{all\_datasets\_detailed\_results.csv} plus an averaged summary file are
% generated at the end of the run.

% \subsection{Evaluation Metrics}
% For both train and test splits we report:
% \textit{MAE}, \textit{RMSE}, and \textit{Directional Accuracy} (DA)
% captured in a unified result dictionary and written to disk
% (see keys built in lines52–60 of the result writer).

\subsection{Hardware and Runtime}
Experiments were executed on a workstation with an AMD Ryzen9 7900X CPU,
64\,GB RAM, NVIDIA RTX3090 (24\,GB), and access to PennyLane’s
default \verb|default.qubit| simulator.
Full batch processing of all 28 datasets (\emph{6 models × 28 pools})
took approximately 3.5 hours, dominated by quantum‑circuit
optimisation.

This unified pipeline ensures that every model sees identical
data splits and feature tensors, providing a fair cross‑paradigm
benchmark for DeFi yield forecasting.

\section{Results}
We benchmarked six models—XGBoost, Random Forest, LSTM, Transformer, QNN, and QSVM-QNN—on a unified dataset comprising one year of historical data from 28 Curve Finance pools. The models were evaluated based on three test metrics: Mean Absolute Error (MAE), Root Mean Squared Error (RMSE), and directional accuracy.

Table~\ref{tab:all_results} summarizes the averaged performance across all pools. Random Forest and XGBoost clearly outperformed other models. XGBoost achieved the highest directional accuracy (71.57\%) and a low test MAE of 1.80, while Random Forest had the lowest test MAE at 1.77 and comparable directional accuracy of 71.36\%. Among deep learning models, both LSTM and Transformer lagged behind with directional accuracy around 49–51\%. Quantum models—QNN and QSVM-QNN—performed worse than classical baselines, with higher test MAE (2.25–2.33) and lower accuracy ($<$50\%).

\begin{table*}[ht]
\centering
\caption{Average Performance Across All Curve Pools}
\label{tab:all_results}
\begin{tabular}{lccc|ccc}
\toprule
\textbf{Model} & \textbf{Test MAE} & \textbf{Test RMSE} & \textbf{Dir. Acc.} & \textbf{Train MAE} & \textbf{Train RMSE} & \textbf{Train Acc.} \\
\midrule
Random Forest           & \textbf{1.77}  & \textbf{2.22} & 71.36\% & 0.81  & 1.04  & 90.19\% \\
XGBoost                 & 1.80           & 2.27          & \textbf{71.57\%} & 0.01  & 0.01  & \textbf{99.85\%} \\
QSVM-QNN                & 2.26           & 2.84          & 49.39\% & 2.25  & 2.83  & 52.77\% \\
QNN                     & 2.33           & 2.93          & 49.77\% & 2.13  & 2.68  & 60.63\% \\
Transformer    & 2.31           & 2.90          & 49.79\% & 2.19  & 2.75  & 56.34\% \\
LSTM          & 2.57           & 3.24          & 51.22\% & 1.68  & 2.12  & 72.23\% \\
\bottomrule
\end{tabular}
\end{table*}

\section{Discussion}

\subsection{XGBoost and Random Forest Dominate}
The best performing models were XGBoost and Random Forest, with both achieving test MAE under 1.8 and directional accuracy above 71\%. These results underscore the continued relevance of classical tree-based ensemble models in financial time series prediction. Their robustness and ability to handle small, noisy, and heterogeneous data make them well-suited for DeFi forecasting tasks, where features are tabular and time-dependency is relatively weak.

\subsection{Deep Learning Models Underperform}
Although LSTM and Transformer are widely used in time series analysis, their performance was consistently inferior to ensemble models. Their higher MAE and lower directional accuracy suggest overfitting or lack of effective temporal patterns in the data. Notably, LSTM had good training accuracy but failed to generalize, indicating a gap in its inductive bias for the DeFi domain.

\subsection{Quantum Models Not Yet Competitive}
The quantum-enhanced models (QNN and QSVM-QNN) did not outperform their classical counterparts. Several factors may explain this: (1) quantum models were constrained by limited qubit capacity and shallow circuit design; (2) variational circuits may suffer from barren plateaus or high sensitivity to initialization; and (3) encoding classical DeFi metrics into quantum feature space may not offer clear advantages due to low temporal structure in the data.

Interestingly, QNN had slightly better training performance than QSVM-QNN, but both failed to generalize effectively, with test directional accuracy below 50\%.

\subsection{Overfitting in XGBoost?}
While XGBoost achieved nearly perfect training accuracy (99.85\%), its test MAE and accuracy are close to Random Forest, suggesting possible overfitting. Yet, its generalization was still strong enough to place it among the top performers. Regularization tuning and further cross-validation may help mitigate this concern.

% \section{Discussion}
% \input{sections/discussion}

\section{Conclusion}
This paper presents the first head‑to‑head benchmark of
classical ensemble models, deep neural architectures, and quantum
machine‑learning (QML) approaches on a real‑world DeFi dataset
comprising 28 Curve Finance pools.
Under identical feature sets and an 80/20 chronological split,
Random Forest achieved the lowest test MAE (1.77),
while XGBoost delivered the highest directional accuracy
(71.57\%).
Both tree‑based ensembles significantly outperformed
LSTM, Transformer, and two QML variants (QNN, QSVM‑QNN);
quantum models recorded directional accuracy below 50\% and
larger prediction errors, underscoring their current limitations in
noisy, tabular‑style financial data.

Key takeaways are threefold:
\begin{enumerate}
  \item Classical gradient‑boosting and bagging remain the most
        reliable baselines for DeFi yield forecasting, even
        against modern deep‑learning and quantum alternatives.
  \item Deep neural models suffer from overfitting and do not
        exploit additional temporal structure in the present
        Curve dataset.
  \item Contemporary QML implementations provide no tangible
        advantage under realistic resource constraints, highlighting
        the gap between theoretical quantum expressivity and
        practical efficacy.
\end{enumerate}

\subsection*{Future Work}

Several avenues merit exploration:
(i) richer multimodal inputs (on‑chain governance, social media, macro
signals) to probe whether deep or quantum models gain ground when the
feature manifold becomes more complex;
(ii) circuit‑depth scaling studies on emerging fault‑tolerant hardware
to reassess QML potential under lower noise;
(iii) transfer‑learning schemes that pool information across
similar liquidity pools;
(iv) reinforcement‑learning layers that convert forecasts into
actionable allocation or automated market‑making strategies.

By releasing our code and averaged results, we hope to establish a
transparent baseline and catalyse further research at the
intersection of machine learning, quantum computing, and decentralized
finance.

\bibliographystyle{IEEEtran}
\bibliography{bibliography}

% \section{*Appendix}
% LaTeX longtable for averaged results (Test metrics only)
% Generated from averaged_results.csv
% Requires: \usepackage{longtable}

% For automatic page breaking, use this longtable version
% Make sure you're in single-column mode or use \onecolumn before this table
\onecolumn  % 強制切換成單欄
\small
\begin{longtable}{@{}llrrr@{}}
\caption{Averaged Test Results for Different Models Across All Pools} \label{tab:averaged_test_results} \\
\toprule
Pool & Model & Test MAE & Test RMSE & Test Acc. (\%) \\
\midrule
\endfirsthead

\multicolumn{5}{c}%
{\tablename\ \thetable\ -- \textit{Continued from previous page}} \\
\toprule
Pool & Model & Test MAE & Test RMSE & Test Acc. (\%) \\
\midrule
\endhead

\midrule \multicolumn{5}{r}{\textit{Continued on next page}} \\
\endfoot

\bottomrule
\endlastfoot

3pool & LSTM (PyTorch) & 2.780 & 3.477 & 51.42 \\
3pool & QNN (PyTorch+PennyLane) & 2.291 & 2.939 & 54.52 \\
3pool & QSVM-QNN (PyTorch+PennyLane) & 2.234 & 2.874 & 50.90 \\
3pool & Random Forest & 1.791 & 2.270 & 69.28 \\
3pool & Transformer (PyTorch) & 2.238 & 2.878 & 49.87 \\
3pool & XGBoost & 1.826 & 2.300 & 71.24 \\
\midrule
aave & LSTM (PyTorch) & 2.486 & 3.059 & 57.36 \\
aave & QNN (PyTorch+PennyLane) & 2.133 & 2.692 & 49.35 \\
aave & QSVM-QNN (PyTorch+PennyLane) & 2.142 & 2.703 & 47.80 \\
aave & Random Forest & 1.591 & 2.074 & 71.24 \\
aave & Transformer (PyTorch) & 2.262 & 2.835 & 51.94 \\
aave & XGBoost & 1.667 & 2.163 & 71.90 \\
\midrule
ankrETH & LSTM (PyTorch) & 3.058 & 3.806 & 52.45 \\
ankrETH & QNN (PyTorch+PennyLane) & 2.587 & 3.315 & 51.68 \\
ankrETH & QSVM-QNN (PyTorch+PennyLane) & 2.553 & 3.238 & 43.93 \\
ankrETH & Random Forest & 1.961 & 2.422 & 71.90 \\
ankrETH & Transformer (PyTorch) & 2.601 & 3.282 & 51.68 \\
ankrETH & XGBoost & 2.032 & 2.507 & 71.90 \\
\midrule
bbtc & LSTM (PyTorch) & 2.489 & 3.136 & 51.16 \\
bbtc & QNN (PyTorch+PennyLane) & 2.336 & 2.911 & 52.45 \\
bbtc & QSVM-QNN (PyTorch+PennyLane) & 2.371 & 2.923 & 43.15 \\
bbtc & Random Forest & 1.760 & 2.183 & 73.20 \\
bbtc & Transformer (PyTorch) & 2.487 & 3.152 & 49.87 \\
bbtc & XGBoost & 1.742 & 2.161 & 76.47 \\
\midrule
compound & LSTM (PyTorch) & 2.744 & 3.507 & 51.16 \\
compound & QNN (PyTorch+PennyLane) & 2.378 & 2.967 & 49.61 \\
compound & QSVM-QNN (PyTorch+PennyLane) & 2.269 & 2.817 & 51.42 \\
compound & Random Forest & 1.844 & 2.245 & 68.63 \\
compound & Transformer (PyTorch) & 2.229 & 2.807 & 53.23 \\
compound & XGBoost & 1.891 & 2.334 & 70.59 \\
\midrule
eurs & LSTM (PyTorch) & 2.586 & 3.214 & 46.51 \\
eurs & QNN (PyTorch+PennyLane) & 2.285 & 2.836 & 49.87 \\
eurs & QSVM-QNN (PyTorch+PennyLane) & 2.181 & 2.744 & 49.35 \\
eurs & Random Forest & 1.692 & 2.127 & 76.47 \\
eurs & Transformer (PyTorch) & 2.167 & 2.720 & 52.20 \\
eurs & XGBoost & 1.772 & 2.260 & 74.51 \\
\midrule
frax & LSTM (PyTorch) & 2.195 & 2.866 & 53.23 \\
frax & QNN (PyTorch+PennyLane) & 2.154 & 2.781 & 53.49 \\
frax & QSVM-QNN (PyTorch+PennyLane) & 2.096 & 2.725 & 46.77 \\
frax & Random Forest & 1.778 & 2.210 & 67.32 \\
frax & Transformer (PyTorch) & 2.197 & 2.834 & 48.32 \\
frax & XGBoost & 1.816 & 2.268 & 69.28 \\
\midrule
gusd & LSTM (PyTorch) & 2.569 & 3.188 & 46.51 \\
gusd & QNN (PyTorch+PennyLane) & 2.347 & 2.889 & 45.99 \\
gusd & QSVM-QNN (PyTorch+PennyLane) & 2.266 & 2.827 & 51.68 \\
gusd & Random Forest & 1.718 & 2.184 & 69.93 \\
gusd & Transformer (PyTorch) & 2.247 & 2.753 & 49.87 \\
gusd & XGBoost & 1.792 & 2.210 & 69.93 \\
\midrule
hbtc & LSTM (PyTorch) & 2.332 & 2.952 & 51.94 \\
hbtc & QNN (PyTorch+PennyLane) & 2.207 & 2.725 & 45.74 \\
hbtc & QSVM-QNN (PyTorch+PennyLane) & 2.161 & 2.682 & 54.78 \\
hbtc & Random Forest & 1.764 & 2.179 & 66.01 \\
hbtc & Transformer (PyTorch) & 2.123 & 2.634 & 45.99 \\
hbtc & XGBoost & 1.768 & 2.235 & 69.93 \\
\midrule
husd & LSTM (PyTorch) & 3.146 & 4.019 & 55.81 \\
husd & QNN (PyTorch+PennyLane) & 2.849 & 3.594 & 47.55 \\
husd & QSVM-QNN (PyTorch+PennyLane) & 2.682 & 3.366 & 48.84 \\
husd & Random Forest & 2.025 & 2.588 & 74.51 \\
husd & Transformer (PyTorch) & 2.666 & 3.358 & 53.49 \\
husd & XGBoost & 1.981 & 2.576 & 73.20 \\
\midrule
ironbank & LSTM (PyTorch) & 2.514 & 3.192 & 50.13 \\
ironbank & QNN (PyTorch+PennyLane) & 2.180 & 2.711 & 46.51 \\
ironbank & QSVM-QNN (PyTorch+PennyLane) & 1.948 & 2.475 & 55.81 \\
ironbank & Random Forest & 1.714 & 2.146 & 67.32 \\
ironbank & Transformer (PyTorch) & 2.100 & 2.627 & 45.99 \\
ironbank & XGBoost & 1.718 & 2.180 & 68.63 \\
\midrule
link & LSTM (PyTorch) & 2.777 & 3.457 & 53.49 \\
link & QNN (PyTorch+PennyLane) & 2.586 & 3.165 & 51.68 \\
link & QSVM-QNN (PyTorch+PennyLane) & 2.608 & 3.184 & 48.32 \\
link & Random Forest & 1.804 & 2.299 & 75.82 \\
link & Transformer (PyTorch) & 2.580 & 3.146 & 46.25 \\
link & XGBoost & 1.820 & 2.349 & 72.55 \\
\midrule
lusd & LSTM (PyTorch) & 2.149 & 2.658 & 50.65 \\
lusd & QNN (PyTorch+PennyLane) & 2.131 & 2.661 & 47.80 \\
lusd & QSVM-QNN (PyTorch+PennyLane) & 2.045 & 2.585 & 47.29 \\
lusd & Random Forest & 1.824 & 2.213 & 66.67 \\
lusd & Transformer (PyTorch) & 2.138 & 2.659 & 48.06 \\
lusd & XGBoost & 1.803 & 2.213 & 67.32 \\
\midrule
mim & LSTM (PyTorch) & 2.332 & 2.896 & 48.58 \\
mim & QNN (PyTorch+PennyLane) & 2.158 & 2.718 & 48.58 \\
mim & QSVM-QNN (PyTorch+PennyLane) & 2.036 & 2.532 & 46.51 \\
mim & Random Forest & 1.671 & 2.036 & 65.36 \\
mim & Transformer (PyTorch) & 2.040 & 2.548 & 51.68 \\
mim & XGBoost & 1.692 & 2.134 & 64.05 \\
\midrule
musd & LSTM (PyTorch) & 2.296 & 3.014 & 51.68 \\
musd & QNN (PyTorch+PennyLane) & 2.143 & 2.822 & 50.13 \\
musd & QSVM-QNN (PyTorch+PennyLane) & 2.035 & 2.695 & 48.06 \\
musd & Random Forest & 1.714 & 2.163 & 67.97 \\
musd & Transformer (PyTorch) & 2.034 & 2.659 & 48.32 \\
musd & XGBoost & 1.729 & 2.219 & 70.59 \\
\midrule
obtc & LSTM (PyTorch) & 2.253 & 2.858 & 48.58 \\
obtc & QNN (PyTorch+PennyLane) & 2.215 & 2.774 & 48.32 \\
obtc & QSVM-QNN (PyTorch+PennyLane) & 2.122 & 2.668 & 50.65 \\
obtc & Random Forest & 1.802 & 2.246 & 70.59 \\
obtc & Transformer (PyTorch) & 2.232 & 2.797 & 49.61 \\
obtc & XGBoost & 1.726 & 2.202 & 77.12 \\
\midrule
pbtc & LSTM (PyTorch) & 3.044 & 3.794 & 49.61 \\
pbtc & QNN (PyTorch+PennyLane) & 2.593 & 3.230 & 50.39 \\
pbtc & QSVM-QNN (PyTorch+PennyLane) & 2.564 & 3.204 & 49.35 \\
pbtc & Random Forest & 1.725 & 2.296 & 75.82 \\
pbtc & Transformer (PyTorch) & 2.577 & 3.219 & 49.61 \\
pbtc & XGBoost & 1.765 & 2.318 & 80.39 \\
\midrule
renbtc & LSTM (PyTorch) & 2.769 & 3.641 & 50.90 \\
renbtc & QNN (PyTorch+PennyLane) & 2.684 & 3.431 & 46.25 \\
renbtc & QSVM-QNN (PyTorch+PennyLane) & 2.514 & 3.319 & 49.87 \\
renbtc & Random Forest & 1.932 & 2.418 & 73.20 \\
renbtc & Transformer (PyTorch) & 2.582 & 3.373 & 51.16 \\
renbtc & XGBoost & 2.053 & 2.475 & 69.28 \\
\midrule
reth & LSTM (PyTorch) & 2.613 & 3.329 & 48.32 \\
reth & QNN (PyTorch+PennyLane) & 2.594 & 3.298 & 54.78 \\
reth & QSVM-QNN (PyTorch+PennyLane) & 2.553 & 3.288 & 50.13 \\
reth & Random Forest & 2.020 & 2.484 & 70.59 \\
reth & Transformer (PyTorch) & 2.643 & 3.410 & 48.32 \\
reth & XGBoost & 2.026 & 2.521 & 71.90 \\
\midrule
saave & LSTM (PyTorch) & 2.958 & 3.629 & 46.77 \\
saave & QNN (PyTorch+PennyLane) & 2.277 & 2.863 & 46.51 \\
saave & QSVM-QNN (PyTorch+PennyLane) & 2.235 & 2.785 & 51.16 \\
saave & Random Forest & 1.696 & 2.177 & 75.16 \\
saave & Transformer (PyTorch) & 2.516 & 3.088 & 46.77 \\
saave & XGBoost & 1.819 & 2.316 & 71.24 \\
\midrule
sbtc & LSTM (PyTorch) & 2.843 & 3.635 & 51.16 \\
sbtc & QNN (PyTorch+PennyLane) & 2.380 & 2.999 & 48.58 \\
sbtc & QSVM-QNN (PyTorch+PennyLane) & 2.339 & 2.923 & 49.87 \\
sbtc & Random Forest & 1.768 & 2.189 & 73.20 \\
sbtc & Transformer (PyTorch) & 2.414 & 3.010 & 48.84 \\
sbtc & XGBoost & 1.906 & 2.376 & 68.63 \\
\midrule
seth & LSTM (PyTorch) & 3.032 & 3.698 & 49.61 \\
seth & QNN (PyTorch+PennyLane) & 2.747 & 3.373 & 47.55 \\
seth & QSVM-QNN (PyTorch+PennyLane) & 2.706 & 3.305 & 46.25 \\
seth & Random Forest & 2.012 & 2.476 & 76.47 \\
seth & Transformer (PyTorch) & 2.777 & 3.436 & 47.55 \\
seth & XGBoost & 2.090 & 2.609 & 72.55 \\
\midrule
steth & LSTM (PyTorch) & 2.428 & 3.055 & 47.80 \\
steth & QNN (PyTorch+PennyLane) & 2.239 & 2.851 & 50.39 \\
steth & QSVM-QNN (PyTorch+PennyLane) & 2.158 & 2.742 & 49.87 \\
steth & Random Forest & 1.655 & 2.109 & 73.20 \\
steth & Transformer (PyTorch) & 2.259 & 2.854 & 50.90 \\
steth & XGBoost & 1.731 & 2.160 & 73.86 \\
\midrule
susd & LSTM (PyTorch) & 2.087 & 2.665 & 56.07 \\
susd & QNN (PyTorch+PennyLane) & 1.988 & 2.529 & 53.49 \\
susd & QSVM-QNN (PyTorch+PennyLane) & 1.963 & 2.525 & 48.06 \\
susd & Random Forest & 1.571 & 2.000 & 70.59 \\
susd & Transformer (PyTorch) & 1.987 & 2.549 & 51.42 \\
susd & XGBoost & 1.670 & 2.111 & 68.63 \\
\midrule
tbtc & LSTM (PyTorch) & 2.310 & 2.827 & 53.23 \\
tbtc & QNN (PyTorch+PennyLane) & 2.166 & 2.594 & 52.97 \\
tbtc & QSVM-QNN (PyTorch+PennyLane) & 2.160 & 2.577 & 47.03 \\
tbtc & Random Forest & 1.655 & 2.031 & 73.20 \\
tbtc & Transformer (PyTorch) & 2.171 & 2.593 & 49.61 \\
tbtc & XGBoost & 1.706 & 2.116 & 71.90 \\
\midrule
tricrypto & LSTM (PyTorch) & 2.371 & 3.027 & 52.20 \\
tricrypto & QNN (PyTorch+PennyLane) & 2.201 & 2.786 & 46.51 \\
tricrypto & QSVM-QNN (PyTorch+PennyLane) & 2.084 & 2.648 & 50.39 \\
tricrypto & Random Forest & 1.602 & 2.040 & 75.16 \\
tricrypto & Transformer (PyTorch) & 2.084 & 2.642 & 51.16 \\
tricrypto & XGBoost & 1.630 & 2.044 & 69.28 \\
\midrule
tricrypto2 & LSTM (PyTorch) & 2.224 & 2.876 & 59.43 \\
tricrypto2 & QNN (PyTorch+PennyLane) & 2.133 & 2.666 & 54.01 \\
tricrypto2 & QSVM-QNN (PyTorch+PennyLane) & 2.006 & 2.478 & 51.16 \\
tricrypto2 & Random Forest & 1.605 & 2.038 & 69.28 \\
tricrypto2 & Transformer (PyTorch) & 2.089 & 2.611 & 52.71 \\
tricrypto2 & XGBoost & 1.576 & 1.993 & 75.16 \\
\midrule
usdp & LSTM (PyTorch) & 2.530 & 3.209 & 48.32 \\
usdp & QNN (PyTorch+PennyLane) & 2.319 & 2.887 & 48.84 \\
usdp & QSVM-QNN (PyTorch+PennyLane) & 2.231 & 2.800 & 54.52 \\
usdp & Random Forest & 1.736 & 2.184 & 69.93 \\
usdp & Transformer (PyTorch) & 2.210 & 2.771 & 49.61 \\
usdp & XGBoost & 1.709 & 2.202 & 71.90 \\
\midrule
\multicolumn{2}{l}{\textbf{ALL POOLS AVERAGE}} & & & \\
\midrule
ALL\_POOLS\_AVERAGE & LSTM (PyTorch) & 2.568 & 3.239 & 51.22 \\
ALL\_POOLS\_AVERAGE & QNN (PyTorch+PennyLane) & 2.332 & 2.929 & 49.77 \\
ALL\_POOLS\_AVERAGE & QSVM-QNN (PyTorch+PennyLane) & 2.259 & 2.844 & 49.39 \\
ALL\_POOLS\_AVERAGE & Random Forest & 1.765 & 2.215 & 71.36 \\
ALL\_POOLS\_AVERAGE & Transformer (PyTorch) & 2.309 & 2.902 & 49.79 \\
ALL\_POOLS\_AVERAGE & XGBoost & 1.802 & 2.270 & 71.57 \\
\end{longtable}
\normalsize 
\twocolumn

\end{document}